%% file: main.tex
\documentclass[conference]{IEEEtran} 
\usepackage{graphicx}
\usepackage{epsfig}
\usepackage{epstopdf}
\usepackage{amsmath,amsthm,amsfonts,amssymb}
\usepackage{multirow}
\usepackage{cite}
\usepackage{tikz}
\usepackage{ellipsis}
\usepackage{xcolor}
\usepackage{xifthen}
\usepackage{cancel}
\theoremstyle{definition} \newtheorem{defi}{Definition}
\newtheorem{prop}{Proposition}
\theoremstyle{definition} 
\theoremstyle{definition} \newtheorem{cor}{Corollary}

\input{figures/commands.tex}
\begin{document}

\title{On Edge Caching with Secrecy Constraints}

\author{\IEEEauthorblockN{Fr\'ed\'eric Gabry, Valerio Bioglio, Ingmar Land}
\IEEEauthorblockA{Mathematical and Algorithmic Sciences Lab\\ France Research Center, Huawei Technologies Co. Ltd.\\
Email: $\{$valerio.bioglio,frederic.gabry,ingmar.land$\}$@huawei.com}} 

\maketitle

\begin{abstract}
In this paper we investigate the problem of optimal cache placement under secrecy constraints in heterogeneous networks, where small-cell base stations are equipped with caches to reduce the overall backhaul load. For two models for eavesdropping attacks, we formally derive the necessary conditions for secrecy and we derive the corresponding achievable backhaul rate. In particular we formulate the optimal caching schemes with secrecy constraints as a convex optimization problem. We then thoroughly investigate the backhaul rate performance of the heterogeneous network with secrecy constraints using numerical simulations. We compare the system performance with and without secrecy constraints and we analyze the influence of the system parameters, such as the file popularity, size of the library files and the capabilities of the small-cell base stations, on the overall performance of our optimal caching strategy. Our results highlight the considerable impact of the secrecy requirements on the overall caching performance of the network.

\end{abstract}

\section{Introduction}
\label{sec:intro}
\input{introduction.tex}

\section{System Model}
\label{sec:model}
\input{system-model.tex}

\section{Secrecy Performance of MDS Coded Caching}
\label{sec:perf}
\input{results.tex}

\section{Numerical Illustrations}
\label{sec:num}
\input{numerical_illustrations.tex}

\section{Conclusions}
\label{sec:conclusions}
\input{future-work.tex}

\bibliographystyle{IEEEbib}
\bibliography{distributed_caching}

\end{document}

%% file: figures/commands.tex
\pgfmathsetseed{1}
\usetikzlibrary{shapes,positioning,arrows,calc,graphs}
\usetikzlibrary{decorations.pathreplacing,decorations.markings,shapes.geometric}
\tikzset{naming/.style={align=center,font=\small}}
\tikzset{antenna/.style={insert path={-- coordinate (ant#1) ++(0,0.25) -- +(135:0.25) + (0,0) -- +(45:0.25)}}}
\tikzset{station/.style={naming,draw,shape=dart,shape border rotate=90, minimum width=10mm, minimum height=10mm,outer sep=0pt,inner sep=3pt}}
\tikzset{mobile/.style={naming,draw,shape=rectangle,minimum width=12mm,minimum height=6mm, outer sep=0pt,inner sep=3pt}}
\tikzset{radiation/.style={{decorate,decoration={expanding waves,angle=90,segment length=4pt}}}}

\newcommand{\MBS}[1]{%
\begin{tikzpicture}
\node[station] (base) {#1};

\draw[line join=bevel] (base.100) -- (base.80) -- (base.110) -- (base.70) -- (base.north west) -- (base.north east);
\draw[line join=bevel] (base.100) -- (base.70) (base.110) -- (base.north east);

\draw[line cap=rect] ([xshift=-.1768cm,yshift=.6pt]base.north -| base.right tail) [antenna=1];
\draw[line cap=rect] ([yshift=.6pt]ant1 |- base.north) -- node[above,shape=rectangle,inner ysep=+.3333em]{\dots} ([xshift=.1768cm,yshift=.6pt]base.north -| base.left tail) [antenna=2];

\end{tikzpicture}
}

\newcommand{\BS}[1]{%
\begin{tikzpicture}
\node[station] (base) {#1};

\draw[line join=bevel] (base.100) -- (base.80) -- (base.110) -- (base.70) -- (base.north west) -- (base.north east);
\draw[line join=bevel] (base.100) -- (base.70) (base.110) -- (base.north east);

\draw[line cap=rect] ([yshift=0pt]base.north) [antenna=1];
\end{tikzpicture}
}

%% file: introduction.tex
Caching content at the wireless  edge, as proposed in \cite{fem_2012}, is a promising technique for future 5G wireless networks \cite{cache_magazine}. The concept of edge caching stems from the idea of significantly reducing the backhaul usage and thus the latency in content retrieval by bringing the content closer to the end users. Exploiting the new capabilities of future multi-tier networks, numerous recent works have investigated the potential benefits of caching data in densely deployed small-cell base stations (SBS) equipped with storage capabilities \cite{proactive,r_1}.

One important perspective to study the performance limits and trade-offs of caching in wireless networks is the information-theoretic perspective, which has gained considerable traction in recent years. In \cite{caching_networks}, caching metrics are defined and analyzed for large networks, while in \cite{fund_lim}, the authors study the fundamental performance limits of caching using network coding techniques. Other approaches have been investigated to analyze caching networks, e.g. in \cite{modeling_tradeoffs} where the problem of optimal placement of cached content is studied in terms of outage probability.  Another particularly important performance measure for edge caching is the overall energy consumption, or equivalently the energy efficiency of the network. Numerous works have investigated the performance of caching from an energy perspective: in \cite{EEICC} it was shown that caching at the edge can provide significant gains in terms of energy efficiency while in \cite{ee_conext}, the authors study the tradeoff between transport and caching energy.

However there exists a performance metric for edge caching networks which has not been studied in the literature yet, namely the \emph{secrecy} metric. Indeed, while secrecy concerns in various wireless networks have been thoroughly investigated in the past decade, see e.g. \cite{sec09} for various topics on secrecy in wireless networks, and while several works have investigated secrecy in heterogeneous networks (HetNets), see e.g. \cite{hetnetssec}, the interest in the secrecy of HetNets with caching has only grown recently. In \cite{limits_cache_sec} secrecy in caching networks is investigated from an information-theoretic perspective using network coding, while in \cite{limits_d2d_sec} the secrecy of device-to-device caching networks is considered.


Departing from these works and building on the model introduced in \cite{globecom}, where the caching strategy was optimized with respect to backhaul load minimization without any secrecy constraint, we consider in this paper a similar model with secrecy requirements. We investigate in this paper two fundamental eavesdropping models during the delivery phase. 
Our main contributions are:

\begin{itemize}
\item We formally define the problem of secure caching at the wireless edge for a HetNet scenario.
\item We derive the secrecy conditions for two eavesdropper models and the corresponding backhaul rate of a caching scheme based on storing Maximum Distance Separable (MDS) encoded packets, and we formulate the optimal caching scheme as a convex optimization problem.
\item We investigate the performance of the optimal secure scheme for a relevant HetNet scenario. 
\item Our results show that the considerable impact of the secrecy constraints on the performance of caching content at the wireless edge.
\end{itemize}
This paper is organized as follows. 
In Section \ref{sec:model}, we define our system model, caching scheme and performance measures. 
In Section \ref{sec:perf}, we derive the main theoretical results of the paper. 
In Section \ref{sec:num} we thoroughly investigate the performance of our optimal schemes and we compare it to other caching schemes in a heterogeneous network scenario. 
Finally, Section~\ref{sec:conclusions} concludes this paper.

\input{figures/geo_topology.tex}

%% file: figures/geo_topology.tex
\begin{figure}[ht!]
 \centering
\resizebox{0.35\textwidth}{!}{  \begin{tikzpicture}
 \clip (-6.7,-6.7) rectangle (6.7,6.7);
    \coordinate (Origin)   at (0,0);
    \coordinate (XAxisMin) at (-5,0);
    \coordinate (XAxisMax) at (5,0);
    \coordinate (YAxisMin) at (0,-5);
    \coordinate (YAxisMax) at (0,5);
   \draw [thick,black] (0,0) circle (6cm);
 
\foreach \x in {1,...,10}{
\foreach \y in {1,...,10}{
\pgfmathparse{rand} \pgfmathsetmacro\xa{\pgfmathresult}
\pgfmathparse{rand}\pgfmathsetmacro\ya{\pgfmathresult} 
\draw [thick,fill=yellow!30] (5*\xa,5*\ya) circle (0.1cm);
}}

\node[blue,thick,inner sep=0pt] (MbS){\MBS{MBS}} (0,0);
    \foreach \x in {-2,...,2}{
      \foreach \y in {-2,...,2}{
       \ifthenelse{\NOT 0 = \x \OR \NOT 0 = \y}{

        \node[scale=0.6] at (2*\x,2*\y) {\BS{SBS}} {};
        \draw[fill=none,dashed,green] (2*\x,2*\y) circle (1.7cm);
  \node  at (2*\x,2*\y) {};}{}
      }
    }   
    \filldraw[fill=gray, fill opacity=0.3, draw=black] (2,2)
        rectangle (4,0);
         \node[scale=0.6] at (0,6) {\BS{SBS}} {};
        \draw[fill=none,dashed,green] (0,6) circle (1.7cm);
  \node  at (0,6) {};
  
    \node[scale=0.6] at (0,-6) {\BS{SBS}} {};
        \draw[fill=none,dashed,green] (0,-6) circle (1.7cm);
  \node  at (0,-6) {};
  
    \node[scale=0.6] at (6,0) {\BS{SBS}} {};
        \draw[fill=none,dashed,green] (6,0) circle (1.7cm);
  \node  at (6,0) {};
  
    \node[scale=0.6] at (-6,0) {\BS{SBS}} {};
        \draw[fill=none,dashed,green] (-6,0) circle (1.7cm);
  \node  at (-6,0) {};
 \end{tikzpicture}}
  \caption{Heterogeneous network.}
  \label{fig:geo_topology}
\end{figure}
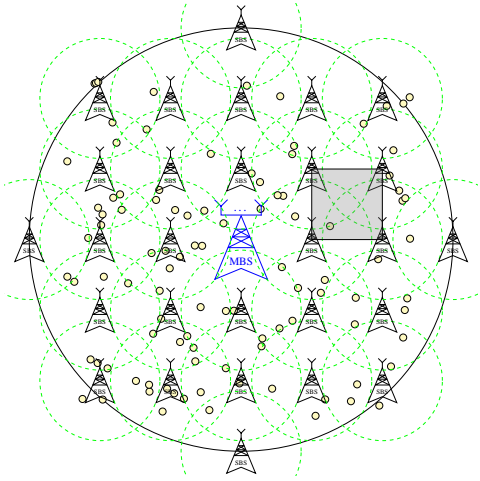

%% file: system-model.tex
In this section, we describe our system model for heterogeneous networks. 
Moreover, we define the MDS encoded caching schemes used in the secrecy problem. 
\subsection{Network Model}
\label{sub:Net}
In the Heterogeneous Network illustrated in Fig.~\ref{fig:geo_topology}, $U$ wireless users request files from a macro-cell base (MBS) station, which has access to a library of $N$ files $\mathcal{F} = \{ F_1, \dots, F_N \}$ of size $B$ bits. 
Users request files stored in the library according to a known popularity distribution, i.e., each file $F_j$ is requested with probability $p_j$, and $\sum_{j=1}^{N} p_j = 1$.
In the following, we denote the probability distribution vector of the files by $p = \{ p_1, \dots, p_N \}$. 


As illustrated in Fig.~\ref{fig:geo_topology}, $N_{\text{SBS}}$ small-cell base stations (SBS) are deployed in the coverage area of the MBS in order to serve user requests within short distance. 
As long as a user is within the coverage range $r$ of the SBS, we assume the content to be delivered without errors. 
Each SBS is equipped with a cache able to store up to $M < N$ complete files, i.e., the cache has size $M \cdot B$ bits.
Each user requesting for files in $\mathcal{F}$ contacts all the SBSs in its coverage range, and it is initially served by them. 
As a consequence, each user $u \in \{ 1, \dots, U \}$ can be served by multiple SBSs, where the number $d_u$ of SBSs serving the user depends on its location (see Fig.~\ref{fig:square}). 
In particular, we call $\gamma_i$ the probability for a user to be served by $d_u = i$ SBSs. 
If the user cannot collect enough information to recover the requested file, the $\text{MBS}$ is contacted to send the missing data to the user, using the backhaul connection. 
The topology of the overall network may vary during time. 
However, at each instant $t$, the connection network can be described as in Fig.~\ref{fig:graph}.

\input{figures/graph.tex}

\subsection{MDS Codes for Caching}
In a MDS($k,n$) codes, $k \geq n$ encoded packets are created such that any subset of $n$ packets are necessary and sufficient to recover the initial information. 
The most known example of MDS codes are the Reed-Solomon (RS) codes. 
From a security point of view, a useful property of these codes is that if less than $n$ packets are received, no part of the encoded information can be retrieved. 
However, RS codes have a high encoding and decoding complexity, which would lead to an increase of the latency due to the decoding of the encoded packets. 
Moreover, the MBS has to store $n$ encoded packets per file, or re-encode the needed packets at each request. 

To solve the complexity problem, rateless codes can be used instead \cite{fount_mckay}. 
Using rateless codes, any packet can be encoded independently on the others, i.e., a new packet can be created on the fly. 
Rateless codes are almost-MDS, i.e., any $(1 + \epsilon) n$ set of the $k$ encoded packets are sufficient to recover the initial information with high probability, with $\epsilon \rightarrow 0$ for $k \rightarrow \infty$. 
Rateless codes have low-complexity encoding and decoding, and new encoded packets can be created by the MBS when a new request is received. 
Moreover, it has be proven that only a negligible part of the information can be retrieved if less than $n$ encoded packets are collected \cite{opt_rateless}. In the remainder of the paper, we will use rateless codes, and it can easily be shown that these codes constitute the worst-case scenario in terms of secrecy performance compared to MDS codes.

\subsection{Coded Caching Scheme}
\label{sub:scheme}
Formally, a caching scheme is constituted of two phases, namely the \emph{placement phase} and the \emph{delivery phase}. 
In the placement phase, which typically occurs at a moment with a low amount of file requests, e.g. at night, the caches of the $\text{SBS}$s are filled according to the placement strategy. 
In the delivery phase, the users send requests to the $\text{MBS}$, and are initially served by the $\text{SBS}$s covering their locations. 
If a user does not receive enough information to recover the requested file, the $\text{MBS}$ is contacted to send the missing data. 

In order to increase the efficiency of the caches, encoded packets are stored instead full files. 
Hence, each file is split into $n$ \emph{fragments}, i.e., $F_j = \{ f_{1}^{(j)}, \cdots, f_{n}^{(j)} \}$ for all $1 \leq j \leq N$. 
Those fragments are used to create $k_j$ \emph{encoded packets} $\{ e_{1}^{(j)}, \cdots, e_{k_j}^{(j)} \}$. 
Each $\text{SBS}$ receives $m_j$ different encoded packets for each file $F_j$ to be stored in its cache, with $\mathbf{m}=[m_{1} \cdots m_{N}]$. 
Formally, the proposed encoded caching scheme is described the following: 
\begin{enumerate}
\item \emph{Placement phase}: The $\text{MBS}$ creates $k_j = N_{SBS} m_j$ encoded packets using a rateless code, sending $m_j$ of them to each SBS. 
\item \emph{Delivery phase}: A user requesting file $F_j$ contacts $d_u \geq 1$ $\text{SBS}$s, receiving $m_j d_u$ disjoint encoded packets. 
If $m_j d_u \geq n$, the user can recover the file due to the MDS-property of rateless codes. 
Otherwise, the $\text{MBS}$ creates and sends the remaining $n - m_j d_u$ encoded packets. 
Due to the MDS-property of rateless codes, the user can recover the requested file, since innovative packets are transmitted with high probability. 
\end{enumerate}

The \emph{secrecy placement problem} consists of finding the optimal number of packets $m_j$ of each file to be stored in the caches in order to minimize the average backhaul rate, which is defined as the average fraction of files that needs to be downloaded from the MBS during the delivery phase, according to secrecy constraints defined in the next section.

\subsection{Optimal Caching with Rateless Codes}
First we briefly review the average backhaul rate of the proposed coded caching scheme without secrecy constraint. 
\begin{prop}[\cite{globecom} Prop. 1]
\label{prop:rate}
The average backhaul rate for an encoded caching placement scheme $\mathcal{C}_{\mathbf{m}}^{\text{MDS}}$ defined by the placement $\mathbf{m}=[m_{1} \cdots m_{N}]$ can be calculated as 
\begin{equation}
\label{eq:th_rate}
R_{(\mathcal{C}_{\mathbf{m}}^{\text{MDS}})} = \sum_{d=1}^{S} \sum_{j=1}^{N} \gamma_d p_j \left( 1 - \min \left( 1, \frac{d m_j}{n} \right) \right),
\end{equation}
where $S \leq N_{\text{SBS}}$ is the maximum number of $\text{SBS}$s serving a user.
\end{prop}

Based on Proposition \ref{prop:rate}, the placement problem of MDS encoded packets can be recast as an optimization problem.
\begin{prop}[\cite{globecom} Prop. 4]
\label{prop:glob}
Finding the optimal MDS coded placement scheme $\mathcal{C}_{(\text{opt})}^{\text{MDS}}$ defined by $\mathbf{m}_{(\text{opt})} = [m_{1} \cdots m_{N}]$, which minimizes the average backhaul rate $R_{(\mathcal{C}_{\mathbf{m}}^{\text{MDS}})}$, is a convex optimization problem:
\begin{align}
\min_{q_1,\dots,q_N} &\sum_{d=1}^{S} \sum_{j=1}^{N} \gamma_d p_j ( 1 - \min ( 1, d q_j ) ) \nonumber \\
\displaystyle 
\text{s.t.} \: &\sum_{j=1}^{N} q_j = M \\
\:\: &0 \leq q_j \leq 1 \qquad \forall j \in [1,N] \nonumber
\end{align}
where we define $q_j \triangleq m_j/n$ in the remainder of the paper.
\end{prop}
 
\input{figures/square.tex}

%% file: figures/graph.tex
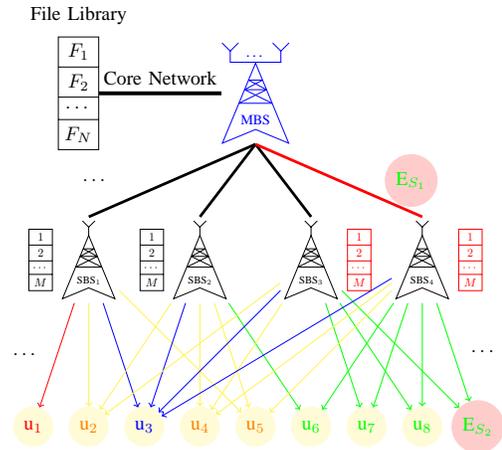
\begin{figure}[!t]
  \centering
\resizebox{0.37\textwidth}{!}{ \begin{tikzpicture}
[stack/.style={rectangle split, rectangle split parts=4, draw, anchor=center},stackR/.style={rectangle split, rectangle split parts=4, draw, red, anchor=center},
  myarrow/.style={single arrow, draw=none}]
 
  [->,>=stealth',shorten >=1pt,auto,
                    semithick]
\node[blue,thick,inner sep=0pt,scale=0.8] (MBS){\MBS{MBS}} (0,2);
\node[stack,left= 2.2cm of MBS] (db) {$F_1$\nodepart{two}$F_2$\nodepart{three}$\cdots$\nodepart{four}$F_N$} (0,0);
\node[above= 0.1cm of db] {File Library};
\path (db.east) edge[thick,line width=0.7mm] node[above] {Core Network} (MBS.west);
\node[scale=0.6]  at (-3,-3) (SBS1) {\BS{SBS$_1$}};
\node[scale=0.6]  at (-1,-3) (SBS2) {\BS{SBS$_2$}};
\node[scale=0.6]  at (1,-3) (SBS3) {\BS{SBS$_3$}};
\node[scale=0.6]  at (3,-3) (SBS4) {\BS{SBS$_4$}};
\node[stack,left= 0.1cm of SBS1,scale=0.6] (db1) {$1$\nodepart{two}$2$\nodepart{three}$\cdots$\nodepart{four}$M$};
\node[stack,left= 0.1cm of SBS2,scale=0.6] (db2) {$1$\nodepart{two}$2$\nodepart{three}$\cdots$\nodepart{four}$M$};
\node[stackR,right= 0.1cm of SBS3,scale=0.6] (db3) {$1$\nodepart{two}$2$\nodepart{three}$\cdots$\nodepart{four}$M$};
\node[stackR,right= 0.1cm of SBS4,scale=0.6] (db4) {$1$\nodepart{two}$2$\nodepart{three}$\cdots$\nodepart{four}$M$};
\node[fill=yellow!20,draw=none,text=red,circle]  at (-4,-6) (u1) {u$_1$};
\node[fill=yellow!20,draw=none,text=orange,circle]  at (-3,-6) (u2) {u$_2$};
\node[fill=yellow!20,draw=none,text=blue,circle]  at (-2,-6) (u3) {u$_3$};
\node[fill=yellow!20,draw=none,text=orange,circle]  at (-1,-6) (u4) {u$_4$};
\node [fill=yellow!20,draw=none,text=orange,circle]  at (0,-6) (u5) {u$_5$};
\node[fill=yellow!20,draw=none,text=green,circle]  at (1,-6) (u6) {u$_6$};
\node[fill=yellow!20,draw=none,text=green,circle]  at (2,-6) (u7) {u$_7$};
\node[fill=yellow!20,draw=none,text=green,circle]  at (3,-6) (u8) {u$_8$};
\node[fill=red!20,draw=none,text=green,circle]  at (4,-6) (u9) {E$_{S_{2}}$};
\path (SBS1.north) edge[thick,line width=0.5mm] node[left=1cm] {$\cdots$} (MBS.south);
\path (SBS2.north) edge[thick,line width=0.5mm] node {} (MBS.south);
\path (SBS3.north) edge[thick,line width=0.5mm] node {} (MBS.south);
\path (SBS4.north) edge[thick,red,line width=0.5mm] node[right=0.8cm,fill=red!20,draw=none,text=green,circle] {E$_{S_{1}}$} (MBS.south);
\path (SBS1) edge[->,red] node[left=0.2cm,text=black] {$\cdots$} (u1);
\path (SBS1) edge[->,yellow!80] node {} (u2);
\path (SBS1) edge[->,yellow!80] node {} (u5);
\path (SBS1) edge[->,blue] node {} (u3);
\path (SBS2) edge[->,yellow!80] node {} (u2);
\path (SBS2) edge[->,blue] node {} (u3);
\path (SBS2) edge[->,yellow!80] node {} (u4);
\path (SBS2) edge[->,yellow!80] node {} (u5);
\path (SBS2) edge[->,green] node {} (u6);
\path (SBS3) edge[->,yellow!80] node {} (u2);
\path (SBS3) edge[->,yellow!80] node {} (u4);
\path (SBS3) edge[->,blue] node {} (u3);
\path (SBS3) edge[->,green] node {} (u7);
\path (SBS3) edge[->,green] node {} (u8);
\path (SBS3) edge[->,green] node {} (u9);
\path (SBS4) edge[->,blue] node {} (u3);
\path (SBS4) edge[->,yellow!80] node {} (u4);
\path (SBS4) edge[->,yellow!80] node {} (u5);
\path (SBS4) edge[->,green] node {} (u6);
\path (SBS4) edge[->,green] node {} (u7);
\path (SBS4) edge[->,green] node {} (u8);
\path (SBS4) edge[->,green] node[right=0.2cm,text=black] {$\cdots$} (u9);
\end{tikzpicture}}
 \caption{Instantaneous HetNet with Secrecy Constraints.}
  \label{fig:graph}
\end{figure}

%% file: figures/square.tex
\def\firstcircle{(1,1) circle (1.7cm)}
\def\secondcircle{(1,-1) circle (1.7cm)}
\def\thirdcircle{(-1,-1) circle (1.7cm)}
\def\fourthcircle{(-1,1) circle (1.7cm)}
\def\rectangle{(-1,1) rectangle (1,-1)}
\begin{figure}[ht!]
  \centering
  \resizebox{0.28\textwidth}{!}{
  \begin{tikzpicture}[scale=2]
  \clip (-1.5,-1.5) rectangle (1.5,1.5);
   \draw[black,fill=blue!20] (-1,1)
        rectangle (1,-1);

    \draw[fill=none,dashed,green] \firstcircle ;
    \draw[fill=none,dashed,green] \secondcircle;
    \draw[fill=none,dashed,green] \thirdcircle ;
 \draw[fill=none,dashed,green] \fourthcircle ;

    \begin{scope}
      \clip \rectangle;
      \fill[red!10] \firstcircle;
    \end{scope}
     \begin{scope}
      \clip \rectangle;
      \fill[red!10] \secondcircle;
    \end{scope}
     \begin{scope}
      \clip \rectangle;
      \fill[red!10] \thirdcircle;
    \end{scope}
    \begin{scope}
      \clip \rectangle;
      \fill[red!10] \fourthcircle;
    \end{scope}

    \begin{scope}
      \clip \rectangle;
      \clip \firstcircle;
      \fill[green!10] \secondcircle;
    \end{scope}

 \begin{scope}
      \clip \rectangle;
      \clip \firstcircle;
      \fill[green!10] \fourthcircle;
    \end{scope}
    
     \begin{scope}
      \clip \rectangle;
      \clip \thirdcircle;
      \fill[green!10] \secondcircle;
    \end{scope}
    
     \begin{scope}
      \clip \rectangle;
      \clip \thirdcircle;
      \fill[green!10] \fourthcircle;
    \end{scope}
    
     \begin{scope}
      \clip \rectangle;
      \clip \secondcircle;
      \clip \thirdcircle;
      \fill[yellow!10] \fourthcircle;
    \end{scope}
    
      \begin{scope}
      \clip \rectangle;
      \clip \secondcircle;
      \clip \thirdcircle;
      \fill[yellow!10] \firstcircle;
    \end{scope}
    
      \begin{scope}
      \clip \rectangle;
      \clip \firstcircle;
      \clip \fourthcircle;
      \fill[yellow!10] \secondcircle;
    \end{scope}
    
        \begin{scope}
      \clip \rectangle;
      \clip \firstcircle;
      \clip \fourthcircle;
      \fill[yellow!10] \thirdcircle;
    \end{scope}
    
          \begin{scope}
      \clip \rectangle;
      \clip \firstcircle;
      \clip \secondcircle;
      \clip \fourthcircle;
      \fill[blue!10] \thirdcircle;
    \end{scope}

\foreach \x in {1,...,3}{
\foreach \y in {1,...,3}{
\pgfmathparse{rand} \pgfmathsetmacro\xa{\pgfmathresult}
\pgfmathparse{rand}\pgfmathsetmacro\ya{\pgfmathresult} 
\draw [thick,fill=yellow!30] (\xa,\ya) circle (0.05cm);
}}
 
   \node[scale=0.9] at (-1,-1) {\BS{SBS$_1$}} {};
        \node[scale=0.9] at (-1,1) {\BS{SBS$_2$}} {};
        \node[scale=0.9] at (1,-1) {\BS{SBS$_3$}} {};
        \node[scale=0.9] at (1,1) {\BS{SBS$_4$}} {};
         \draw[fill=none,dashed,green] \firstcircle ;
    \draw[fill=none,dashed,green] \secondcircle;
    \draw[fill=none,dashed,green] \thirdcircle ;
 \draw[fill=none,dashed,green] \fourthcircle ;
   \end{tikzpicture}}
  \caption{Small cells topology.}
  \label{fig:square}
\end{figure}
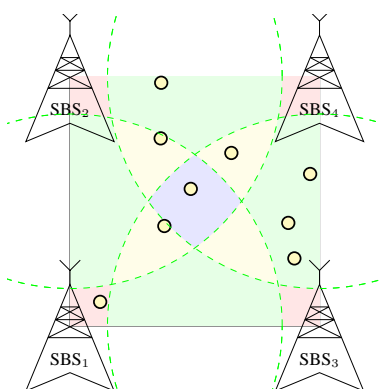

%% file: results.tex
In this section we first describe the two models of eavesdropping attacks investigated in this paper. Then, we derive the secrecy conditions for both eavesdropper scenarios S$_1$ and S$_2$. Moreover we derive for each scenario the achievable backhaul rate as a convex optimization problem.


We should first note that we restrict ourselves to eavesdropping attacks on the backhaul links and caches, i.e., we do not consider the potential eavesdropping of wireless communications from the small-cell base stations to the mobile users. This choice is justified by the fact that caching strategies are independent from this eavesdropping model which is inherent to the broadcast nature of wireless communications and has been studied extensively for networks without caching, see e.g. \cite{sec09} and references therein. Hence in the following we consider two different types of attackers compared to usual wireless eavesdroppers. Both scenarios are depicted in Figure \ref{fig:graph}.
First, we define the secrecy condition for the caching heterogeneous network.
\begin{defi}
\label{def:1}
The caching network is said to be secure for a scenario S$_i$ if for any eavesdropper $E_{S_i}$ in that scenario, i.e. for any MBS-to-SBS link in Scenario S$_1$ and for any  location of $E_{S_2}$ in Scenario S$_2$, $E_{S_i}$ cannot recover \emph{any full file} $F_j$, $\forall j$.
\end{defi}
We should note that both MDS and rateless codes fulfill Definition \ref{def:1} if fewer than n packets of a file are collected by an eavesdropper. 

\subsection{Backhaul Eavesdropper - Scenario S$_1$}
In Scenario S$_1$ we consider an eavesdropper $E_{S_1}$ which can eavesdrop a single MBS-to-SBS link during the delivery phase. In other words, any packet delivered by the MBS through the corrupted link to the SBS is intercepted by $E_{S_1}$.

For this scenario, we derive the secrecy condition and corresponding achievable backhaul rate as follows. 
Here we denote by $Q$ the number of requests per SBS during the delivery phase. 
Moreover, we assume that each SBS receives the same amount of file requests during the delivery phase (i.e., a geographic uniformity of requests). 
As a consequence, the overall number of requests is given by $\bar{Q} = Q N_{\text{SBS}}$. 


\begin{prop}
\label{prop:backhaul-eve}
The caching network is secure in the presence of an eavesdropper E$_{S_1}$ if and only if 
\begin{equation}
\label{eq:backhaul-eve}
m_j > \frac{n}{Q p_j} \cdot \frac{Q p_j \sum_{d=1}^{I} \gamma_d - 1}{\sum_{d=1}^{I} d \gamma_d}, \: \: j = 1,\dots,N,
\end{equation}
where $I \triangleq \min \left( S, \lfloor n/m_j \rfloor \right)$.
\begin{proof}
If we denote by $Q_j = Q p_j$ the number of requests for file $F_j$ received by the SBS, an eavesdropper E$_{S_1}$, i.e., on a single MBS-SBS link, can collect an amount of packets which depends on the number of SBSs serving each request. 
We denote by $P$ that amount in the remainder of the proof.
With rateless codes, these packets are all different, hence $P<n$ is necessary to prevent the recovery of file k by E$_{S_1}$. 
Moreover, each request results in the transmission of $n( 1 - \min ( 1, d m_j/n ) )$ packets over the eavesdropped link with probability $\gamma_d$. 
Hence, for $j = 1,\dots,N$, we have that $P = \sum_{d=1}^{S} Q \gamma_d p_j n( 1 - \min ( 1, d m_j/n ) )$. 
In order to find a bound for $m_j$, we notice that $\min ( 1, d m_j/n ) ) = 1$ if $d > n/m_j$, i.e., $( 1 - \min ( 1, d m_j/n ) ) = 0$ if $d > n/m_j$. 
Hence the condition $P<n$ can be rewritten, for $j = 1,\dots,N$, as
$\sum_{d=1}^{I} Q \gamma_d p_j n( 1 - d m_j/n ) < n$,
where $I = \min \left( S, \lfloor n/m_j \rfloor \right)$. 
The result \eqref{eq:backhaul-eve} follows from standard calculations.
\end{proof}
\end{prop}

In order to preserve the convexity of the optimization problem, we focus on the worst-case attack scenario where users can contact only one SBS, i.e., $\gamma_1 = 1$.

\begin{cor}
\label{cor:backhaul-eve}
The caching network is secure in the presence of an eavesdropper E$_{S_1}$ if  
\begin{equation}
m_j > n(1 - 1/(Q p_j)), \: \: j = 1,\dots,N.
\end{equation}
\end{cor}

Combining Prop.~\ref{prop:glob} and Cor.~\ref{cor:backhaul-eve} we can derive the optimization problem for Scenario S$_1$. 

\begin{prop}
\label{prop:opt_1}
Finding the optimal coded placement scheme $\mathcal{C}_{(\text{opt})}$ in Scenario S$_1$ defined by $\mathbf{m}_{(\text{opt})} = [m_{1} \cdots m_{N}]$, which minimizes the average backhaul rate $R_{(\mathcal{C}_{\mathbf{m}})}$, is a convex optimization problem:
\begin{align}
\min_{q_1,\dots,q_N} &\sum_{d=1}^{S} \sum_{j=1}^{N} \gamma_d p_j ( 1 - \min ( 1, d q_j ) )\nonumber \\
\displaystyle 
\text{s.t.} \: &\sum_{j=1}^{N} q_j = M \\
\:\: &\left( 1-\frac{1}{Q p_j} \right)^{+} < q_j \leq 1 \qquad \forall j \in [1,N] \nonumber
\end{align}
where $q_j = m_j/n$ and $(\cdot)^{+} = \max(\cdot,0)$.
\end{prop}

According to Cor.~\ref{cor:backhaul-eve}, it is possible to get a lower bound for the cache size.

\begin{cor}
\label{cor:threshold}
The cache size should satisfy the following condition in order to guarantee secrecy in Scenario S$_1$:
\begin{equation}
M > N - \frac{\eta}{Q},
\end{equation}
where $\eta \triangleq \sum_{j=1}^{N} 1/p_j$.
\begin{proof}
Summing the N inequalities $m_j > n(1 - 1/(Q p_j))$ over all the $N$ files, we get $M > N - 1 / Q\sum_{j=1}^{N} p_j $, and the corollary is obtained.
\end{proof}
\end{cor}
Intuitively, Cor.\ref{cor:threshold} reflects the fact that with sufficient cache size, fewer transmissions over the backhaul are needed, and hence the system is better protected against an attacker E$_{S_1}$ able to wiretap the backhaul link. On the other hand, as we will see in the following section, if the eavesdropper E$_{S_2}$ is instead able to access the cache contents, it should be guaranteed that not too many packets of the files are present in each cache.

\subsection{Cache Eavesdropper - Scenario S$_2$}
In Scenario S$_2$ we consider an eavesdropper $E_{S_2}$ which can eavesdrop the content of the caches within coverage range. Hence if the malicious mobile user $E_{S_2}$ can be served by $d_{E_{S_2}}$ SBSs, then $E_{S_2}$ has access to the cache content of these $d_{E_{S_2}}$ corrupted SBSs.

As in the previous section for Scenario S$_1$, we derive the secrecy condition and corresponding achievable backhaul rate for this scenario.

\begin{prop}
\label{prop:cache-eve}
The caching network is secure in the presence of an eavesdropper E$_{S_2}$ if and only if
\begin{equation}
\label{eq_cache_eve}
 m_j<\frac{n}{S}, \:\: j = 1,\dots,N.
 \end{equation}
\begin{proof}
The worst-case eavesdropper E$_{S_1}$ is within coverage of $S$ SBSs, from which it can collect $m_j S$ encoded packets of file $F_j$. 
Since these $m_j S$ packets are all different for the proposed caching scheme, the file $F_j$ can only be if $m_j S \geq n$.
\end{proof}
\end{prop}

By summing Equation \eqref{eq_cache_eve} over the $N$ files we obtain an upper bound on the secure cache size.

\begin{cor}
\label{cor:cache_eve}
The cache size should satisfy the following condition in order to guarantee secrecy in Scenario S$_2$:
\begin{equation}
M < N/S.
\end{equation}
\end{cor}

As mentioned previously, we observe that Cor.~\ref{cor:cache_eve} implies that the cache size is upper-bounded to guarantee secrecy against an eavesdropper E$_{S_2}$ with access to the caches while on the other hand Cor.~\ref{cor:threshold} suggests that the cache size should be higher than a certain threshold to guarantee secrecy against a backhaul wiretapper E$_{S_2}$. These observations will be further analyzed in the numerical illustrations of Section \ref{sec:num}.

Finally from Prop.~\ref{prop:cache-eve} and Prop.~\ref{prop:glob} we can derive the optimization problem for Scenario S$_2$. 

\begin{prop}
\label{prop:opt_2}
Finding the optimal coded placement scheme $\mathcal{C}_{(\text{opt})}$ in Scenario S$_2$ defined by $\mathbf{m}_{(\text{opt})} = [m_{1} \cdots m_{N}]$, which minimizes the average backhaul rate $R_{(\mathcal{C}_{\mathbf{m}})}$, is a convex optimization problem:
\begin{align}
\min_{q_1,\dots,q_N} &\sum_{d=1}^{S} \sum_{j=1}^{N} \gamma_d p_j ( 1 - \min ( 1, d q_j ) ) \nonumber \\
\displaystyle 
\text{s.t.} \: &\sum_{j=1}^{N} q_j = M \\
\:\: &0 \leq q_j < 1/S \qquad \forall j \in [1,N]. \nonumber
\end{align}
\end{prop}

%% file: numerical_illustrations.tex
In this section we investigate the performance of the proposed caching scheme with secrecy constraints in terms of backhaul rate in a HetNet topology of particular interest. 
The presented numerical results can be generalized to any network topology. 

\subsection{Network Topology}
\label{num:topology}
In the simulations, we consider the network illustrated in Fig.~\ref{fig:geo_topology}, where the MBS has a coverage area $\mathcal{R}$ of radius $D=500$ meters. 
$N_\text{SBS} = 316$ SBSs are deployed in a regular grid, with horizontal distance equal to $d=60$ meters. 
We call $r$, with $ d/\sqrt{2} \leq r \leq d$, the radius of the coverage area of each SBS.
As a consequence, the coverage areas of neighbor SBSs may overlap, as shown in the highlighted square of Fig.~\ref{fig:square}. 
We calculate the probability a user has to be served by $d$ SBSs as
\begin{equation*}
\gamma_d = \frac{\rho_d \mathcal{A}_d}{\sum_{i=1}^{S} \rho_i \mathcal{A}_i},
\end{equation*}
where $\mathcal{A}_d$ is the total area of $\mathcal{R}$ where a user can be served by $d$ $\text{SBS}$s and $\rho_d$ is its average density. 
In the following, we consider a uniform distribution of users over $\mathcal{R}$, with density $\rho_d = \rho=0.05$ users$/m^{2}$. 
This corresponds to $U=39,269$ mobile users present in the coverage area of the MBS. 
Finally, the users request files according to a Zipf law of parameter $\alpha$, i.e.,
\begin{equation*}
p_{j}=\frac{1/j^{\alpha}}{\sum_j 1/j^{\alpha}},
\end{equation*}
where $\alpha$ represents the skewness of the distribution  \cite{zipf}. 


\subsection{Numerical Results}

In the following, we study the performance of the proposed optimal cache placement with secrecy constraints. Unless specified otherwise, the SBSs have a coverage area of $r=60$ meters with a storage capacity of $M=20$ files. 
The library contains $N=200$ files, whose popularity follows a Zipf distribution of parameter $\alpha=0.7$, and have $Q=100$ requests. 

\begin{figure}[!t]
\centering
\vspace{-2ex}
\includegraphics[scale=0.46]{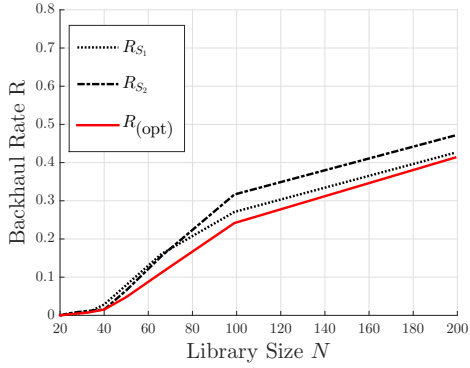}
\vspace{-2ex}
\caption{Backhaul rate as a function of the library size with and without secrecy constraints and with $\alpha = 0.7$, $r=60$, $M=20$, and $Q=100$.}
\vspace{-2ex}
\label{fig:N}
\end{figure}

In Fig.~\ref{fig:N} the backhaul rates $R$ obtained for the eavesdropping scenarios S$_1$ and S$_2$ are compared with the optimal caching placement without secrecy constraints as a function of the size $N$ of the library of files. 
As expected, the backhaul rate grows with the library size $N$, since a smaller library size means an higher cache hit probability. 
For small library sizes, in the cache eavesdropper scenario it is possible to better exploit the cache to decrease the rate, while for bigger values of $N$ the backhaul eavesdropper scenario converges to the optimal one without secrecy. 
This fact can be intuitively explained as follows: when $N$ gets large, there are so many files in the library that the probability that $Q p_j(n-m_j) < n$ tends to 1 as $p_j \rightarrow 0$ and hence the new constraint due to Scenario S$_1$ $\left( 1-\frac{1}{Q p_j} \right)^{+} < q_j$ reduces to the constraint without secrecy $0 < q_j$.
For Scenario S$_2$ the new constraint induced by the secrecy condition does not converge to the case without secrecy when $N$ gets larger, and therefore there is a gap in terms of achievable backhaul rate.

\begin{figure}[!t]
\centering
\vspace{-2ex}
\includegraphics[scale=0.46]{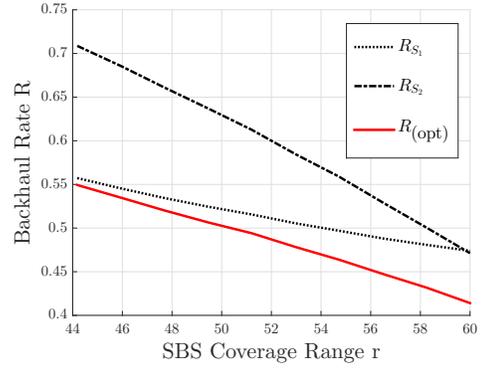}
\vspace{-2ex}
\caption{Backhaul rate as a function of the coverage range with and without secrecy constraints and with $\alpha = 0.7$, $N=200$, $M=20$, and $Q=100$.}
\vspace{-2ex}
\label{fig:R}
\end{figure}

In Fig.~\ref{fig:R} the backhaul rates $R$ of the proposed solutions are calculated as a function of the range $r$ of the coverage of the SBSs. 
With a small coverage area, the SBSs have to store less data in order to defend their caches from $E_{S_2}$, so the backhaul rate is high. 
However, an increase of coverage area means that each cache can store fewer packets of each file, which favors the optimization of the cache eavesdropper secure problem. 
This is due to the fact that the additional constraint in this scenario is an upper bound on the number of encoded packets stored per file. 
A larger coverage area means that more users can contact a large number of SBSs, and hence each SBS needs to store fewer encoded packets of popular files, using the cache to store a small number of packets of less popular ones. 
On the contrary, $E_{S_1}$ can take advantage of the increase of the coverage range, which corresponds to an augmentation of the proportion of the files that are transmitted over the backhaul. 
In this case, the additional constraint is a lower bound, hence it is not possible to store fewer packets of popular files, and the cache is not filled. 
However, the performance increases slightly, since more files reach the lower bound. 

\begin{figure}[!t]
\centering
\includegraphics[scale=0.46]{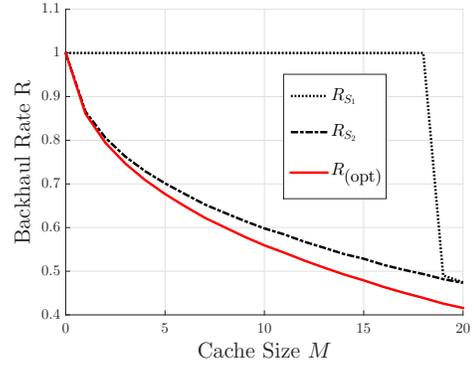}
\vspace{-2ex}
\caption{Backhaul rate as a function of the cache size with and without secrecy constraints S$_1$ and S$_2$ and with $\alpha = 0.7$, $r=60$, $N=200$, and $Q=100$.}
\vspace{-2ex}
\label{fig:M}
\end{figure}

In Fig.~\ref{fig:M} we depict the backhaul rates $R$ as a function of the cache size $M$ for both scenarios and without secrecy. We observe that an increase of $M$ improves the performance in Scenario S$_2$, since a larger cache size available means a larger number of different files stored. 
Since the cache eavesdropper E$_{S_2}$ needs $n$ packets to recover the data, if a small number of encoded packets of less popular files are stored, E$_{S_2}$ cannot recover anything. 
As a consequence, the difference from the optimal caching scheme is due to the less popular files, which are not stored. 
In practice, the SBSs store more packets of popular files, which is sub-optimal from the backhaul rate point of view, and do not store packets of less popular file. This strategy is sub-optimal, but performs only slightly worse than the optimal scheme. 
For Scenario $S_1$, the cache is too small to ensure security, so it is not used.  As a consequence, the backhaul rate remains 1, since the SBSs do not store any packet. 
However, when the cache reaches the critical size, the performance improves since the cache is filled. 
Thus, caching in this scenario is useful only if a certain cache size is available, as stated in Cor.~\ref{cor:threshold}.

\begin{figure}[!t]
\centering
\includegraphics[scale=0.46]{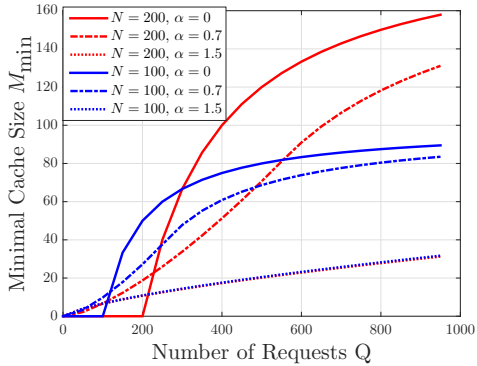}
\vspace{-2ex}
\caption{Minimum cache size for Scenario S$_1$ as a function of the number of requests per SBS $Q$, with $\alpha = 0.7$ and $r=60$.}
\vspace{-2ex}
\label{fig:T}
\end{figure}

Finally, in Fig.~\ref{fig:T} we show the minimal required cache size for guaranteeing secrecy for Scenario S$_1$ as a function of the number of requests $Q$. The blue lines depict the case $N=100$ while the red lines depict $N=200$. As expected, a larger cache size is required for secrecy against a backhaul eavesdropper when N gets larger. Moreover we observe that the minimal cache size $M_{\min}$ is an increasing function of $Q$, and that a smaller Zipf parameter $\alpha$ leads to a larger threshold $M_{\min}$. In other term, if the popularity profile of the files becomes uniform, i.e., $\alpha \rightarrow 0$ it becomes more difficult to ensure secrecy. There is, however, an exception to that behavior, namely for $Q < N$. In this case, the security constraint of Prop.\ref{prop:backhaul-eve} is verified for any M, in particular $M = 0$ already guarantees secrecy since the eavesdropper E$_{S_1}$ does not intercept enough backhaul transmissions to decode a file.

%% file: future-work.tex
We considered the problem of optimal MDS-encoded content placement at the cache-equipped small-cell base stations at the wireless edge \emph{with secrecy constraints}. We defined two types of security threats in the caching network for which we derived the secrecy conditions and the corresponding optimal caching placement strategy based on rateless encoding. We then thoroughly studied the secrecy performance of this optimal placement for a relevant heterogeneous scenario scenario by comparing it to the optimal placement without secrecy concerns and by measuring the influence of the key parameters, such as the capabilities of the small-cell base stations and the library statistics. Our numerical observations, which can easily be generalized to any geometric topology and number of eavesdroppers, highlighted the impact of secrecy constraints on cached content and showed the behavior of the system depending on the type of attack. Further the crucial importance of optimizing securely content placement was highlighted since such an optimization yields a significant decrease of the load of the network backhaul, i.e., a considerable decrease of delay for secure delivery to the end users.